\def\prl{Phys. Rev. Lett.}
\def\prd{Phys. Rev. D}
\def\plb{Phys. Lett. B}
\def\nima{Nucl. Instr. and Meth. A}

\def\Bs{\ensuremath{B_s^0}}
\def\Bsbar{\ensuremath{\bar{B}_s^0}}

\def\Bsst{\ensuremath{B_s^*}}
\def\Bsstbar{\ensuremath{\bar{B}_s^*}}

\def\Bsall{\ensuremath{B_s^{(*)}}}
\def\Bsallbar{\ensuremath{\bar{B}_s^{(*)}}}

\def\BsBs{\ensuremath{\Bs\Bsbar}}
\def\BsstBs{\ensuremath{\Bsst\Bsbar}}
\def\BsstBsst{\ensuremath{\Bsst\Bsstbar}}
\def\BsallBsall{\ensuremath{\Bsall\Bsallbar}}

\def\UfiveS{\ensuremath{\Upsilon(5S)}}
\def\UfourS{\ensuremath{\Upsilon(4S)}}

\def\Bstophig{\ensuremath{\Bs \to \phi \gamma}}
\def\Bstogg{\ensuremath{\Bs \to \gamma \gamma}}

\def\phitoKK{\ensuremath{\phi \to K^+ K^-}}

\def\BztoKstg{\ensuremath{B^{0}\to K^*(892)^{0} \gamma}}
\def\BptoKstg{\ensuremath{B^{+}\to K^*(892)^{+} \gamma}}
\def\BtoKstg{\ensuremath{B \to K^*(892) \gamma}}

\def\BstoDspi{\ensuremath{\Bs \to D_s^- \pi^+}}

\def\BztoKsteta{\ensuremath{B^{0} \to K^*(892)^{0} \eta}}
\def\Bstophieta{\ensuremath{\Bs \to \phi \eta(\gamma\gamma)}}

\def\BtoXsg{\ensuremath{B \to X_s \gamma}}

\def\sample{23.6}

\def\ifb{\ensuremath{\text{fb}^{-1}}}

\def\mevct{\ensuremath{\text{MeV}/c^{2}}}
\def\gevct{\ensuremath{\text{GeV}/c^{2}}}
\def\gev{\text{GeV}}

\def\BF{\ensuremath{{\cal B}}}

\def\mbc{\ensuremath{M_{\mathrm{bc}}}}
\def\deltae{\ensuremath{{\Delta}E}}
\def\SFW{\ensuremath{\rm SFW}}
\def\thhel{\ensuremath{\theta_{\rm hel}}}
\def\costhhel{\ensuremath{\cos{\thhel}}}
\def\costthhel{\ensuremath{\cos^2{\thhel}}}

\def\Ebeam{\ensuremath{E_{\rm beam}^{\rm CM}}}

\def\yieldBsstBsst{\ensuremath{S_{\BsstBsst}}}
\def\yieldBsstBs{\ensuremath{S_{\BsstBs}}}
\def\yieldBsBs{\ensuremath{S_{\BsBs}}}

\def\fBsstBsst{\ensuremath{f_{\BsstBsst}}}

\def\NBs{\ensuremath{N_{\Bs}}}

\def\sigmabb{\ensuremath{\sigma_{b\bar{b}}^{\UfiveS}}}

\def\Lint{\ensuremath{L_{\rm int}}}

\def\yieldBsstBsstBstophig{\ensuremath{18 {^{+6}_{-5}}}}
\def\yieldBsstBsBstophig{\ensuremath{0.5 {^{+2.9}_{-1.9}}}}
\def\yieldBsBsBstophig{\ensuremath{-0.7 {^{+2.5}_{-1.6}}}}

\def\bfBstophig{\ensuremath{(57 { ^{+18}_{-15}  } { ^{+12}_{-11}}) \times 10^{-6} }}
\def\bfBstophigss{\ensuremath{(57 { ^{+18}_{-15}(\rm stat) } { ^{+12}_{-11} (\rm syst) } ) \times 10^{-6} }}
\def\bfBstophigsu{\ensuremath{57 { ^{+18}_{-15}  } { ^{+12}_{-11}} }}

\def\bffsfststBstophig{\ensuremath{(10.3 { ^{+3.2}_{-2.8}  } { \pm 1.3}) \times 10^{-6} }}

\def\systsignifbfBstophig{\ensuremath{5.5}}

\def\effBstophig{\ensuremath{24.7}}

\def\yieldBsstBsstBstogg{\ensuremath{-7.3 {^{+2.4}_{-2.0} } }}
\def\yieldBsstBsBstogg{\ensuremath{-0.8 {^{+4.8}_{-3.8}}}}
\def\yieldBsBsBstogg{\ensuremath{-4.7 {^{+3.9}_{-2.8}} }}

\def\limitbfBstogg{\ensuremath{8.7 \times 10^{-6}}}
\def\limitbfBstoggsu{\ensuremath{8.7}}

\def\effBstogg{\ensuremath{17.8}}

\documentclass[aps,prl,twocolumn,superscriptaddress,showpacs,preprintnumbers,amsmath,amssymb]{revtex4}

\usepackage{graphicx} 
\usepackage{dcolumn}  

\usepackage{color}

\graphicspath{{ps}}

\begin{document}

\preprint{\vbox{ \hbox{   }
                 \hbox{BELLE Preprint 2007-48}
                 \hbox{KEK Preprint 2007-63}
                 \hbox{   }
}}

\title{ \boldmath Observation of $B_s^0 \to \phi \gamma$ and Search for $B_s^0 \to \gamma \gamma$ Decays at Belle}

\affiliation{Budker Institute of Nuclear Physics, Novosibirsk}
\affiliation{Chiba University, Chiba}
\affiliation{University of Cincinnati, Cincinnati, Ohio 45221}
\affiliation{Justus-Liebig-Universit\"at Gie\ss{}en, Gie\ss{}en}
\affiliation{The Graduate University for Advanced Studies, Hayama}
\affiliation{Hanyang University, Seoul}
\affiliation{University of Hawaii, Honolulu, Hawaii 96822}
\affiliation{High Energy Accelerator Research Organization (KEK), Tsukuba}
\affiliation{Institute of High Energy Physics, Chinese Academy of Sciences, Beijing}
\affiliation{Institute of High Energy Physics, Vienna}
\affiliation{Institute of High Energy Physics, Protvino}
\affiliation{Institute for Theoretical and Experimental Physics, Moscow}
\affiliation{J. Stefan Institute, Ljubljana}
\affiliation{Kanagawa University, Yokohama}
\affiliation{Korea University, Seoul}
\affiliation{Kyungpook National University, Taegu}
\affiliation{\'Ecole Polytechnique F\'ed\'erale de Lausanne (EPFL), Lausanne}
\affiliation{Faculty of Mathematics and Physics, University of Ljubljana, Ljubljana}
\affiliation{University of Maribor, Maribor}
\affiliation{University of Melbourne, School of Physics, Victoria 3010}
\affiliation{Nagoya University, Nagoya}
\affiliation{Nara Women's University, Nara}
\affiliation{National Central University, Chung-li}
\affiliation{National United University, Miao Li}
\affiliation{Department of Physics, National Taiwan University, Taipei}
\affiliation{H. Niewodniczanski Institute of Nuclear Physics, Krakow}
\affiliation{Nippon Dental University, Niigata}
\affiliation{Niigata University, Niigata}
\affiliation{University of Nova Gorica, Nova Gorica}
\affiliation{Osaka City University, Osaka}
\affiliation{Osaka University, Osaka}
\affiliation{Panjab University, Chandigarh}
\affiliation{Saga University, Saga}
\affiliation{University of Science and Technology of China, Hefei}
\affiliation{Seoul National University, Seoul}
\affiliation{Sungkyunkwan University, Suwon}
\affiliation{University of Sydney, Sydney, New South Wales}
\affiliation{Toho University, Funabashi}
\affiliation{Tohoku Gakuin University, Tagajo}
\affiliation{Department of Physics, University of Tokyo, Tokyo}
\affiliation{Tokyo Institute of Technology, Tokyo}
\affiliation{Tokyo Metropolitan University, Tokyo}
\affiliation{Tokyo University of Agriculture and Technology, Tokyo}
\affiliation{Virginia Polytechnic Institute and State University, Blacksburg, Virginia 24061}
\affiliation{Yonsei University, Seoul}
  \author{J.~Wicht}\affiliation{\'Ecole Polytechnique F\'ed\'erale de Lausanne (EPFL), Lausanne} 
  \author{I.~Adachi}\affiliation{High Energy Accelerator Research Organization (KEK), Tsukuba} 
  \author{H.~Aihara}\affiliation{Department of Physics, University of Tokyo, Tokyo} 
  \author{K.~Arinstein}\affiliation{Budker Institute of Nuclear Physics, Novosibirsk} 
  \author{V.~Aulchenko}\affiliation{Budker Institute of Nuclear Physics, Novosibirsk} 
  \author{T.~Aushev}\affiliation{\'Ecole Polytechnique F\'ed\'erale de Lausanne (EPFL), Lausanne}\affiliation{Institute for Theoretical and Experimental Physics, Moscow} 
  \author{A.~M.~Bakich}\affiliation{University of Sydney, Sydney, New South Wales} 
  \author{V.~Balagura}\affiliation{Institute for Theoretical and Experimental Physics, Moscow} 
  \author{A.~Bay}\affiliation{\'Ecole Polytechnique F\'ed\'erale de Lausanne (EPFL), Lausanne} 
  \author{K.~Belous}\affiliation{Institute of High Energy Physics, Protvino} 
% ...
\collaboration{The Belle Collaboration}

\noaffiliation

\begin{abstract}
We search for the radiative penguin decays $B_s^0 \to \phi \gamma$ and $B_s^0 \to \gamma \gamma$ in a 23.6 $\text{fb}^{-1}$ data sample collected at the $\Upsilon(5S)$ resonance with the Belle detector at the KEKB $e^+e^-$ asymmetric-energy collider.  We observe for the first time a radiative penguin decay of the $B_s^0$ meson in the $B_s^0 \to \phi \gamma$ mode and we measure ${\cal B}(B_s^0 \to \phi \gamma) = (57 { ^{+18}_{-15}(\rm stat) } { ^{+12}_{-11} (\rm syst) } ) \times 10^{-6}$. No significant $B_s^0 \to \gamma \gamma$ signal is observed and we set a 90\% confidence level upper limit of ${\cal B}(B_s^0 \to \gamma \gamma) < 8.7 \times 10^{-6}$.
\end{abstract}


\maketitle

\tighten

{\renewcommand{\thefootnote}{\fnsymbol{footnote}}}
\setcounter{footnote}{0}

\begin{figure}
\centering
\begin{tabular}{ccc} 
   \includegraphics[width=0.2\textwidth]{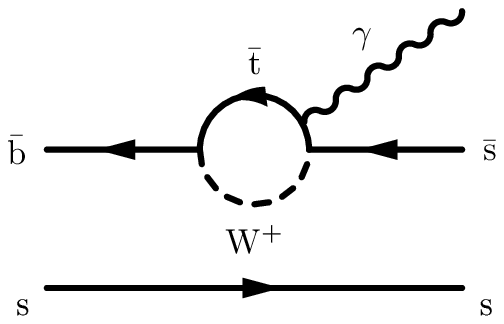} &
   \hspace{0.8cm} &
   \includegraphics[width=0.2\textwidth]{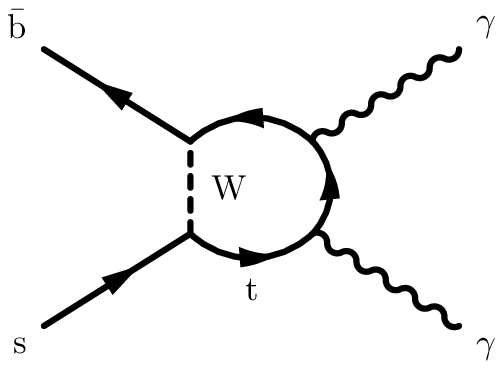} \\
\end{tabular}
\caption{Diagrams describing the dominant processes for the \Bstophig\ (left) and \Bstogg\ (right) decays.}
\label{figure:feynman}
\end{figure}

Radiative penguin decays, which produce a photon via a one-loop Feynman diagram, are a good tool to search for physics beyond the Standard Model (SM) because particles not yet produced in the laboratory can make large contributions to such loop effects. The \Bstophig~\cite{chargeconj} mode is a radiative process described within the SM by a $\bar{b} \to \bar{s} \gamma$ penguin diagram (Fig.~\ref{figure:feynman} left); it is the strange counterpart of the \BtoKstg\ decay, whose observation by CLEO in 1993~\cite{b2kstg-cleo} unambiguously demonstrated the existence of penguin processes. In the SM, the \Bstophig\ branching fraction has been computed with $\sim$$30 \%$ uncertainty to be about $40 \times 10^{-6}$~\cite{bs2phigam-sm1,bs2phigam-sm2}. The \Bstogg\ mode is usually described by a penguin annihiliation diagram (Fig.~\ref{figure:feynman} right), and its branching fraction has been calculated in the SM to be in the range $(0.5-1.0) \times 10^{-6}$~\cite{bs2gamgam-sm1,bs2gamgam-sm2,bs2gamgam-sm3}. Neither \Bstophig\ nor \Bstogg\ has yet been observed, and the upper limits at the 90\% confidence level (CL) on their branching fractions are, respectively, $120 \times 10^{-6}$~\cite{bs2phigam-cdf} and $53 \times 10^{-6}$~\cite{u5s-excl}.

A strong theoretical constraint on the \Bstophig\ branching fraction is generally assumed due to good agreement between SM expectations and experimental results for $b \to s \gamma$ rates, such as in \BptoKstg\ and \BztoKstg\ decays~\cite{bs2phigam-sm1,bs2phigam-sm2,b2kstg-th,PDG2007} or inclusive $B \to X_s \gamma$ decays~\cite{b2xsg-th,PDG2007}. The \Bstogg\ decay rate is constrained in a similar way~\cite{bs2gamgam-xsg}, though various New Physics (NP) scenarios such as supersymmetry with broken $R$-parity~\cite{bs2gamgam-supersym}, a fourth quark generation~\cite{bs2gamgam-4thquark} or a two Higgs doublet model with flavor changing neutral currents~\cite{bs2gamgam-2higgsdoublet}, can increase the \Bstogg\ branching fraction by up to an order of magnitude without violating constraints on the \BtoXsg\ branching fraction.

In this study, we use a data sample with an integrated luminosity (\Lint) of \sample\ \ifb\ that was collected with the Belle detector at the KEKB asymmetric-energy $e^+e^-$ (3.6 on 8.2~\gev) collider~\cite{KEKB} operating at the \UfiveS\ resonance (10.87~\gev).

The Belle detector is a large-solid-angle magnetic spectrometer that
consists of a 4-layer silicon detector (SVD~\cite{SVD2}), a central
drift chamber (CDC), an array of aerogel threshold Cherenkov counters
(ACC), a barrel-like arrangement of time-of-flight scintillation
counters (TOF), and an electromagnetic calorimeter comprised of
CsI(Tl) crystals (ECL) located inside a superconducting solenoid coil
that provides a 1.5~T magnetic field.  An iron flux-return located
outside the coil is instrumented to detect $K_L^0$ mesons and to
identify muons. The detector is described in detail
elsewhere~\cite{Belle}.

The variety of hadronic events at the \UfiveS\ resonance is richer than at the \UfourS. $B^+$, $B^0$ and \Bs\ mesons are all produced in \UfiveS\ decay.  \Bs\ mesons are produced mainly via $\UfiveS \to \BsstBsst$ decays, with subsequent \Bsst\ low energy photon de-excitation.  The $b\bar{b}$ production cross section at the \UfiveS, the fraction of \BsallBsall\ events in the $b\bar{b}$ events, and the fraction of  \BsstBsst\ events among \BsallBsall\ events have been measured to be, respectively, $\sigmabb = (0.302 \pm 0.015)$ nb~\cite{u5s-incl}, $f_s = (19.5^{+3.0}_{-2.3})\%$~\cite{PDG2007} and $\fBsstBsst = (93^{+7}_{-9})\%$~\cite{u5s-excl}. The \BsstBs\ and \BsBs\ decay fractions are small and not yet measured.

Charged tracks are reconstructed using the SVD and CDC detectors and are required to originate from the interaction point. Kaon candidates are selected from charged tracks with the requirement $\mathcal{L}_K/(\mathcal{L}_K+\mathcal{L}_\pi)>0.6$, where $\mathcal{L}_{K}$ ($\mathcal{L}_{\pi}$) is the likelihood for a track to be a kaon (pion) based on the response of the ACC and on measurements from the CDC and TOF. For the selected kaons, the identification efficiency is about 85\% with about 9\% of pions misidentified as kaons.

We reconstruct $\phi$ mesons in the decay mode \phitoKK\ by combining oppositely charged kaons having an invariant mass within $\pm 12$ \mevct\ ($\sim$$2.5\sigma$) of the nominal $\phi$ mass~\cite{PDG2007}.

We reject photons from $\pi^0$ and $\eta$ decays to two photons using a likelihood based on the energy and polar angles of the photons in the laboratory frame and the invariant mass of the photon pair. To reject merged photons from $\pi^0$ decays and neutral hadrons such as neutrons and $K_L^0$, we require an ECL shower shape consistent with that of a single photon: for each cluster, the ratio of the energy deposited in the central $3 \times 3$ calorimeter cells to that of the larger $5 \times 5$ array of cells has to be greater than 0.95. Candidate photons are required to have a signal timing consistent with originating from the same event. For the \Bstogg\ mode, photons are selected in the barrel part of the ECL ($33^{\circ} < \theta < 128^{\circ}$) and we require that the total energy of the event be less than 12 \gev.

\Bs\ meson candidates are selected using the beam-energy-constrained mass $\mbc = \sqrt{(\Ebeam)^2 - (p_{\Bs}^{\rm CM})^2}$ and the energy difference $\deltae = E_{\Bs}^{\rm CM} - \Ebeam$.  In these definitions, \Ebeam\ is the beam energy and $p_{\Bs}^{\rm CM}$ and $E_{\Bs}^{\rm CM}$ are the momentum and the energy of the \Bs\ meson, with all variables being evaluated in the center-of-mass (CM) frame.  We select $\Bs$ meson candidates with $\mbc > 5.3\;\gevct$ for both modes, and $-0.4\;\gev < \deltae < 0.4\;\gev$ for the \Bstophig\ mode and $-0.7\;\gev < \deltae < 0.4\;\gev$ for the \Bstogg\ mode. No events with multiple \Bs\ candidates are observed in either data or Monte Carlo (MC) simulation. \Bsst\ mesons are not fully reconstructed due to the low energy of the photon from the \Bsst\ decay. Signal candidates coming from \BsstBsst, \BsstBs\ and \BsBs\ are well-separated in \mbc, but they overlap in \deltae~\cite{u5s-excl}.

The main background in both search modes is due to continuum events coming from light-quark pair production ($u\bar{u}$, $d\bar{d}$, $s\bar{s}$ and $c\bar{c}$). Rejection of this background is studied and optimized using large signal MC samples and a continuum MC sample having about three times the size of the data sample.  A Fisher discriminant based on modified Fox-Wolfram moments (\SFW~\cite{SFW}) is used to separate signal from continuum background. The process $e^+e^- \to q \bar{q} \gamma$ is a source of high-energy photons with low polar angles and can thus be a background for radiative $B$ decays. Therefore, for the \Bstophig\ mode, we apply a more restrictive \SFW\ requirement when the candidate photon is recontructed outside the barrel part of the ECL. This procedure is not used for the \Bstogg\ mode where photons are selected only in the barrel.  For the \Bstophig\ mode, the \SFW\ requirement is chosen in order to maximize a figure of merit defined as $N_{\rm sig}/\sqrt{N_{\rm sig}+N_{udsc}}$, where $N_{\rm sig}$ and $N_{udsc}$ are the expected number of signal events coming from \BsstBsst\ events and continuum events, respectively. $N_{\rm sig}$ and $N_{udsc}$ are computed in the \Bstophig\ signal window ($\mbc > 5.4~\gevct$, $-0.2 \; \gev < \deltae < 0.02 \; \gev$ and $|\costhhel| < 0.8$) and are normalized to an integrated luminosity of \sample~\ifb\ assuming $\BF(\Bstophig) = 40 \times 10^{-6}$. The helicity angle \thhel\ is the angle between the \Bs\ and the $K^+$ in the $\phi$ rest frame. For signal events \costhhel\ should follow a $1-\costthhel$ distribution, while for continuum events the distribution is found to be flat.  For the \Bstogg\ mode, we optimize the \SFW\ requirement to minimize the 90\% CL upper limit on the branching fraction computed by the Feldman-Cousins method~\cite{feldmancousins}. The upper limit calculation requires two inputs: the number of observed events ($N_{\rm obs}$) and the expected number of background events ($N_{\rm bkg}$). We assume $N_{\rm obs} \equiv N_{\rm sig} + N_{udsc}$ and $N_{\rm bkg} \equiv N_{udsc}$. $N_{\rm sig}$ and $N_{udsc}$ are computed in the \Bstogg\ signal window ($\mbc > 5.4~\gevct$ and $-0.3 \; \gev < \deltae < 0.05 \; \gev$) assuming that $\BF(\Bstogg) = 1.0 \times 10^{-6}$.

Inclusive $b\bar{b}$ backgrounds from \UfiveS\ decays are studied using MC samples having about the same size as the data sample. Backgrounds coming from $B^{+}$ or $B^{0}$ decays are found to lie outside of the fit region. For \Bs\ decays, no event is reconstructed in the \Bstogg\ mode. The \Bstophieta\ decay is a potential background for the \Bstophig\ mode and is studied using a dedicated MC sample. Assuming that its branching fraction is the same as its $B^0$ counterpart \BztoKsteta~\cite{PDG2007}, we expect to reconstruct one \Bstophieta\ background event. Considering the large \Bstophieta\ branching fraction uncertainty, this background is treated as a source of systematic error.

For the \Bstophig\ (\Bstogg) mode, we perform a three-dimensional (two-dimensional) unbinned extended maximum likelihood fit to \mbc, \deltae\ and \costhhel\ (\mbc\ and \deltae) using the probability density functions (PDF) described below.

The signal PDFs for \mbc\ and \deltae\ are modeled separately for events coming from \BsstBsst, \BsstBs\ and \BsBs\ with smoothed two-dimensional histograms built from signal MC events. The \mbc\ (\deltae) mean for the \BsstBsst\ signal is adjusted to the \Bsst\ mass (the \Bsst-\Bs\ mass difference) obtained from \BstoDspi\ events reconstructed in the same \UfiveS\ data sample. The \mbc\ and \deltae\ resolutions for the \Bstophig\ (\Bstogg) signal are corrected using a control sample of \BztoKstg\ events ($e^+e^- \to \gamma \gamma$ events) recorded on the $\UfourS$ resonance. Statistical uncertainties contained in these corrections are included in the systematic uncertainty. Continuum background is modeled with an ARGUS function~\cite{argus} for \mbc\ and a first-order polynomial function for \deltae. For the \Bstophig\ mode, the signal (continuum) PDF for \costhhel\ is modeled with a $1-\costthhel$ (constant) function. The \Bstophieta\ background PDF is modeled using MC events as the product of a two-dimensional PDF for \mbc\ and \deltae\ and a one-dimensional histogram for \costhhel. The likelihood is defined as
\begin{eqnarray}
\mathcal{L} = e^{-\sum_{j}{S_j}} \times \prod_{i}(\sum_{j}{S_j P_j^i}) \,,
\end{eqnarray}
where $i$ runs over all events, $j$ runs over the possible event categories (signals or backgrounds), $S_j$ is the number of events in each category and $P_j$ is the corresponding PDF.

Both fits have six free fit variables: the yields for the \BsstBsst, \BsstBs\ and \BsBs\ signals (\yieldBsstBsst, \yieldBsstBs\ and \yieldBsBs), the continuum background normalization and PDF parameters, except the ARGUS endpoint which is fixed to $5.435~\gev$. The branching fractions ($\BF(\Bstophig)$ and $\BF(\Bstogg)$) are determined from the \BsstBsst\ signal yields with the relations
\begin{eqnarray}
\yieldBsstBsst^{\Bstogg}     & = & \BF(\Bstogg) \times \epsilon_{\gamma\gamma} \times \NBs \times \fBsstBsst \,, \\ \nonumber 
\yieldBsstBsst^{\Bstophig} & = & \BF(\Bstophig) \times \BF(\phitoKK) \\ &  & ~ \times \epsilon_{\phi\gamma} \times \NBs \times \fBsstBsst \,,
\end{eqnarray}
where $\epsilon$'s are the MC signal efficiencies listed in Table~\ref{table:results} and \NBs\ is the number of \Bs\ mesons evaluated as $\NBs = 2 \times \Lint \times \sigmabb \times f_s = (2.8 { ^{+0.5}_{-0.4}}) \times 10^6$. 

In the \Bstophig\ mode we observe \yieldBsstBsstBstophig\ signal events in the \BsstBsst\ region and no significant signals in the two other regions. These signal yields are compatible with $\fBsstBsst = (93^{+7}_{-9})\%$~\cite{u5s-excl}. We measure $\BF(\Bstophig) \times f_s \times \fBsstBsst =  \bffsfststBstophig$ and $\BF(\Bstophig) = \bfBstophig$ with a significance of $5.5\sigma$, where the first uncertainty is statistical and the second is systematic. Systematic uncertainties and computation of the significance are detailed below. The measured branching fraction is in agreement with SM expectations~\cite{bs2phigam-sm1,bs2phigam-sm2} and with the measurements $\BF(\BztoKstg) = (40.1 \pm 2.0) \times 10^{-6}$ and $\BF(\BptoKstg) = (40.3 \pm 2.6) \times 10^{-6}$~\cite{PDG2007}. We observe no significant \Bstogg\ signal and, including systematic uncertainties, determine a 90\% CL upper limit of $\BF(\Bstogg) < \limitbfBstogg$. This limit is about six times more restrictive than the previous one~\cite{u5s-excl}, though still about one order of magnitude larger than SM expectations~\cite{bs2gamgam-sm1,bs2gamgam-sm2,bs2gamgam-sm3} and still above the predictions of NP models~\cite{bs2gamgam-supersym,bs2gamgam-4thquark,bs2gamgam-2higgsdoublet}. The results are summarized in Table~\ref{table:results} and fit projections in the signal windows are shown in Figs.~\ref{figure:data_bs2phig} and~\ref{figure:data_bs2gg}.

\renewcommand{\arraystretch}{1.4}
\begin{table}
\centering
\caption{Efficiencies, signal yields, branching fractions and significances (Sig.) obtained from the fits described in the text. The first uncertainty is statistical and the second systematic. The upper limit is calculated at the 90\% CL.}
\begin{tabular}{l  c  c  c  c c c }\hline
Mode           & $\epsilon$ (\%) & \yieldBsBs         & \yieldBsstBs                      & \yieldBsstBsst         & \BF\ ($10^{-6}$)    & Sig.\ \\ \hline 
$\phi\gamma$   & \effBstophig    & \yieldBsBsBstophig & $\phantom{-}\yieldBsstBsBstophig$ & \yieldBsstBsstBstophig & \bfBstophigsu       & \systsignifbfBstophig \\ 
$\gamma\gamma$ & \effBstogg      & \yieldBsBsBstogg   & \yieldBsstBsBstogg                & \yieldBsstBsstBstogg & $ < \limitbfBstoggsu$ & -- \\ \hline
\end{tabular}
\label{table:results}
\end{table}

\begin{figure}
\centering
 \begin{tabular}{cc}
   \includegraphics[width=0.23\textwidth]{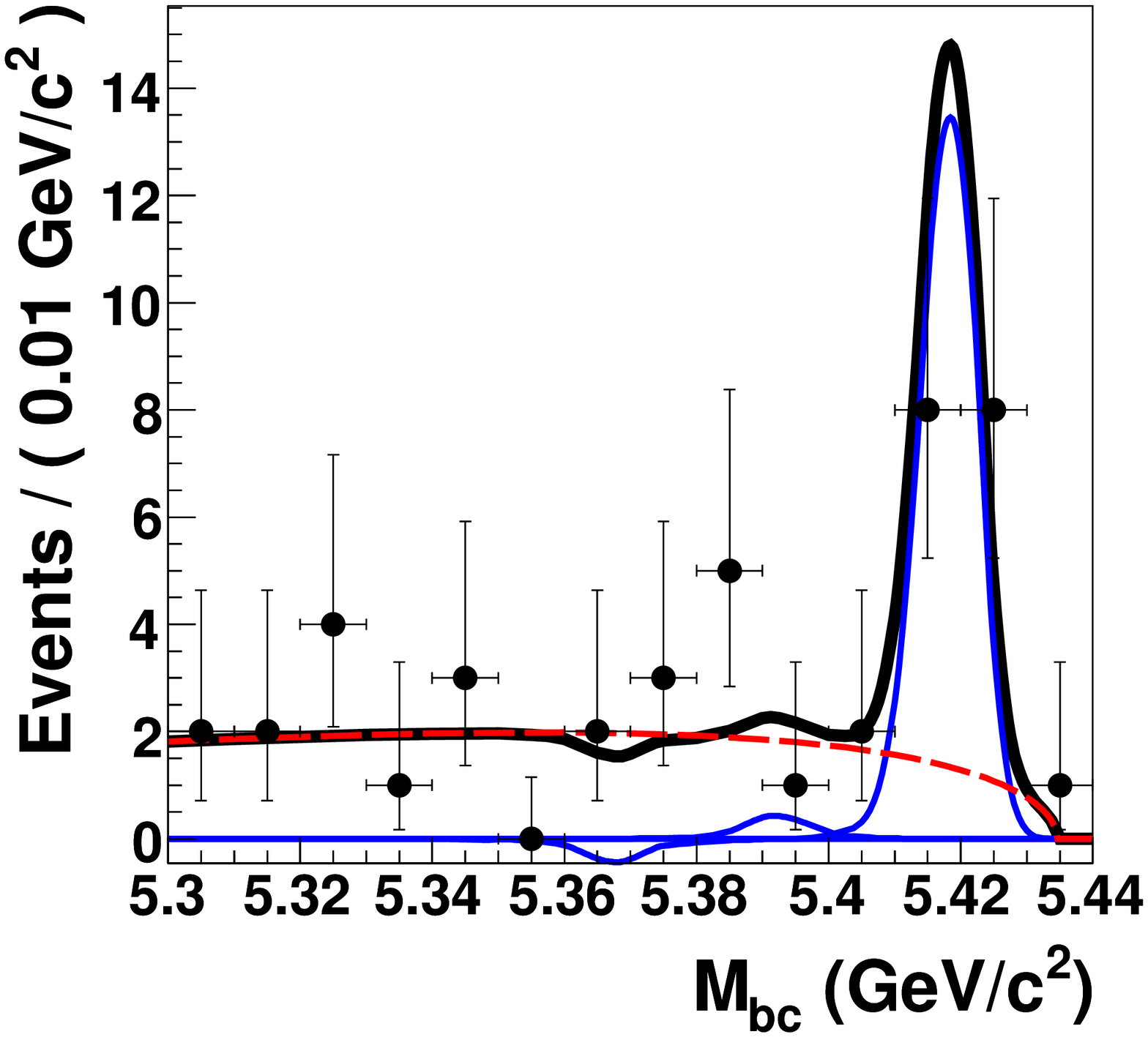} & 
   \includegraphics[width=0.23\textwidth]{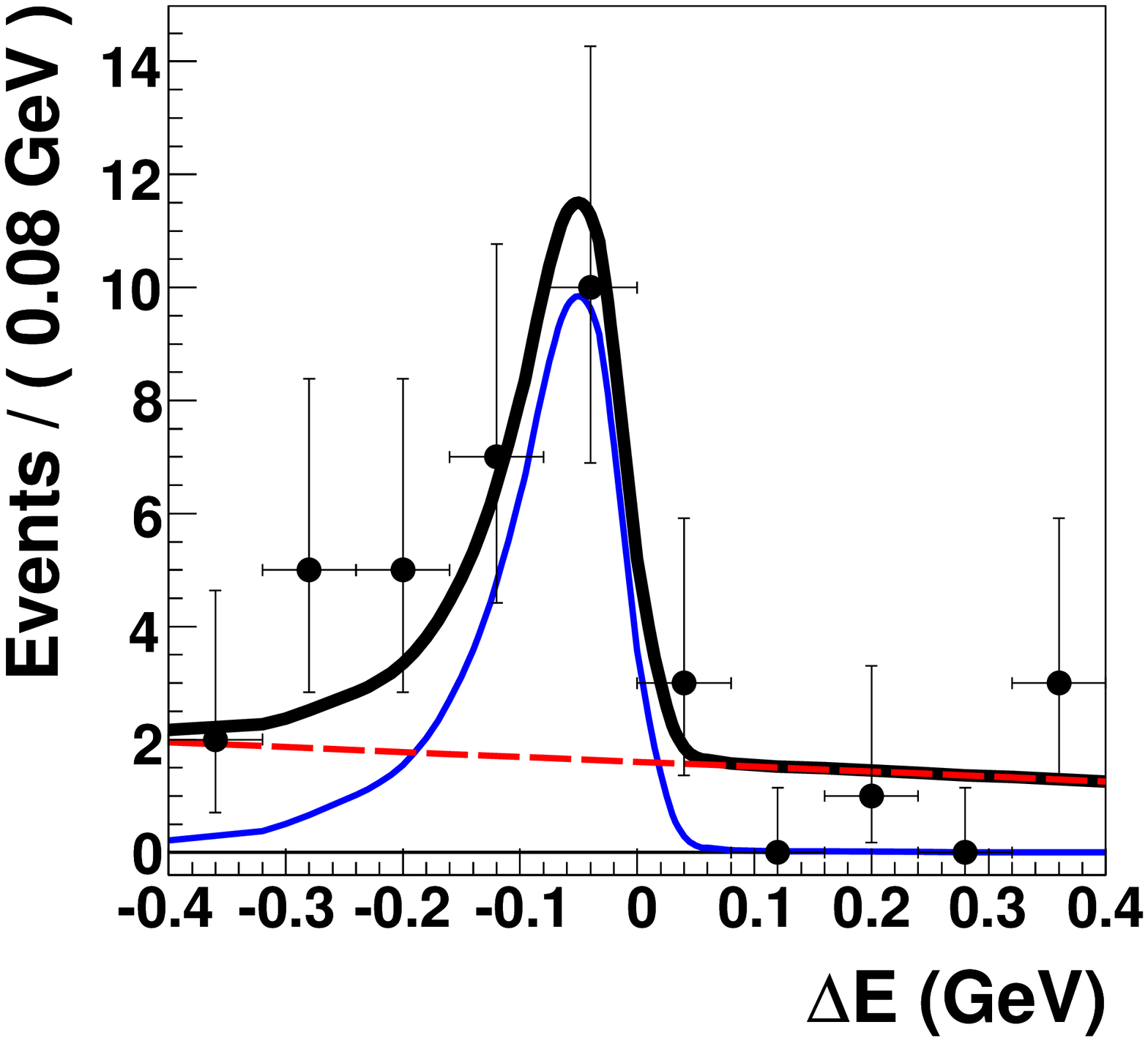} \\
   \includegraphics[width=0.23\textwidth]{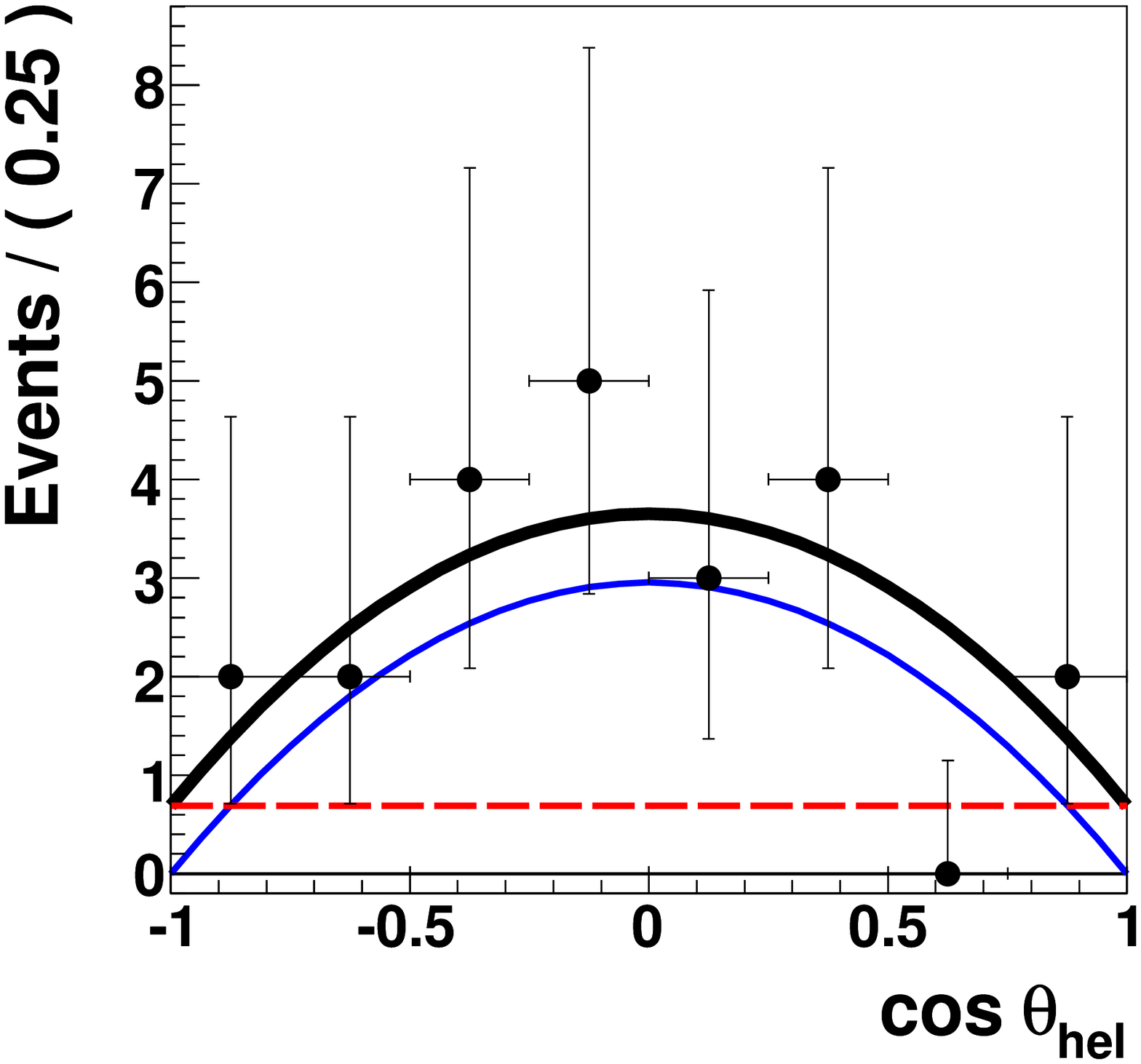} & 
   \includegraphics[width=0.23\textwidth]{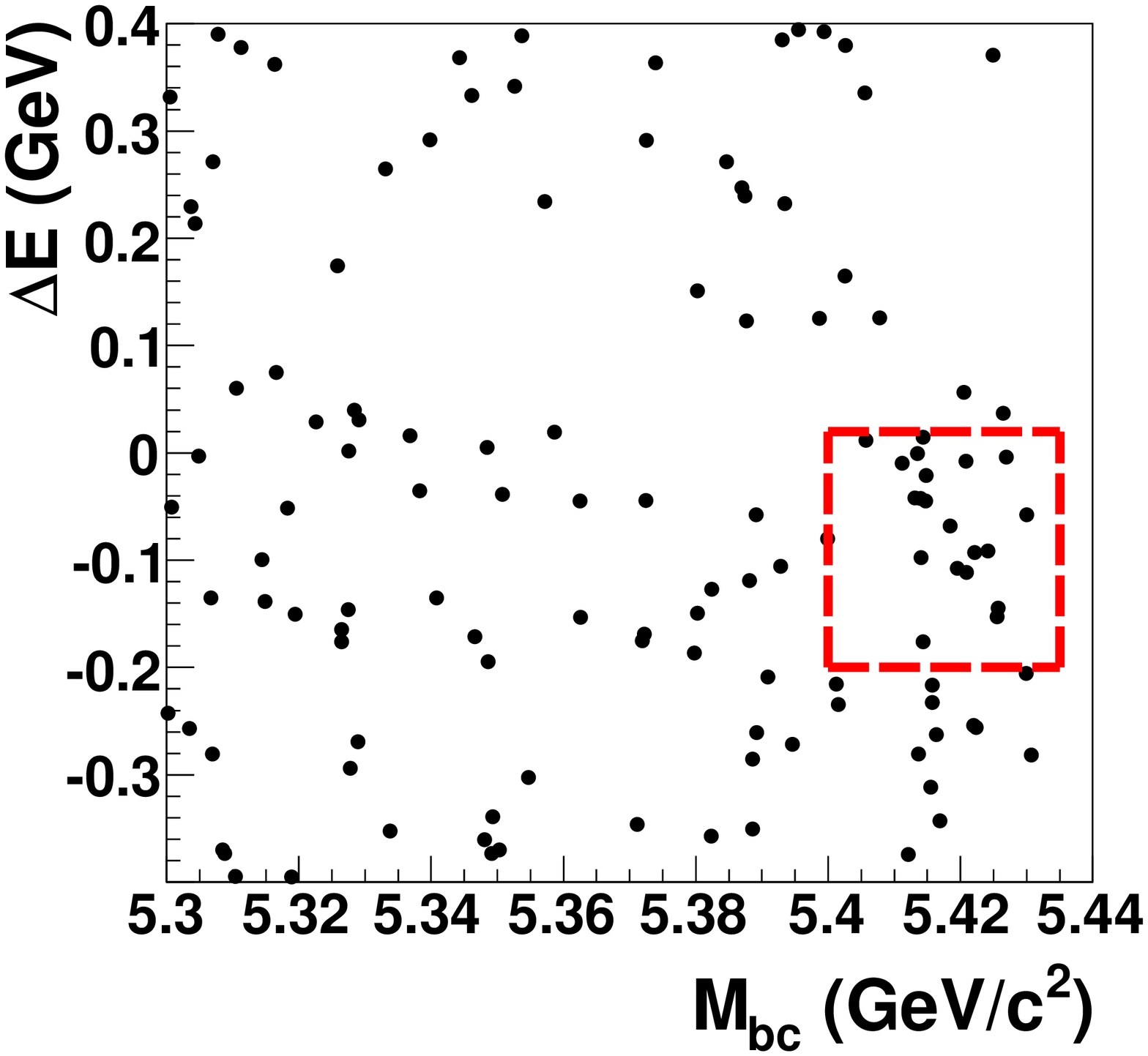} \\
 \end{tabular}
\caption{\mbc, \deltae\ and \costhhel\ projections together with fit results for the \Bstophig\ mode. The points with error bars represent data, the thick solid curves are the fit functions, the thin solid curves are the signal functions, and the dashed curves show the continuum contribution. On the \mbc\ figure, signals from \BsBs, \BsstBs\ and \BsstBsst\ appear from left to right. On the \deltae\ and \costhhel\ figures, due to the requirement $\mbc > 5.4$ \gevct\ only the \BsstBsst\ signal contributes. The bottom right figure shows \deltae\ versus \mbc\ for selected data events. The dashed lines show the signal window.}
\label{figure:data_bs2phig}
\end{figure}

\begin{figure}
\centering
 \begin{tabular}{cc}
   \includegraphics[width=0.23\textwidth]{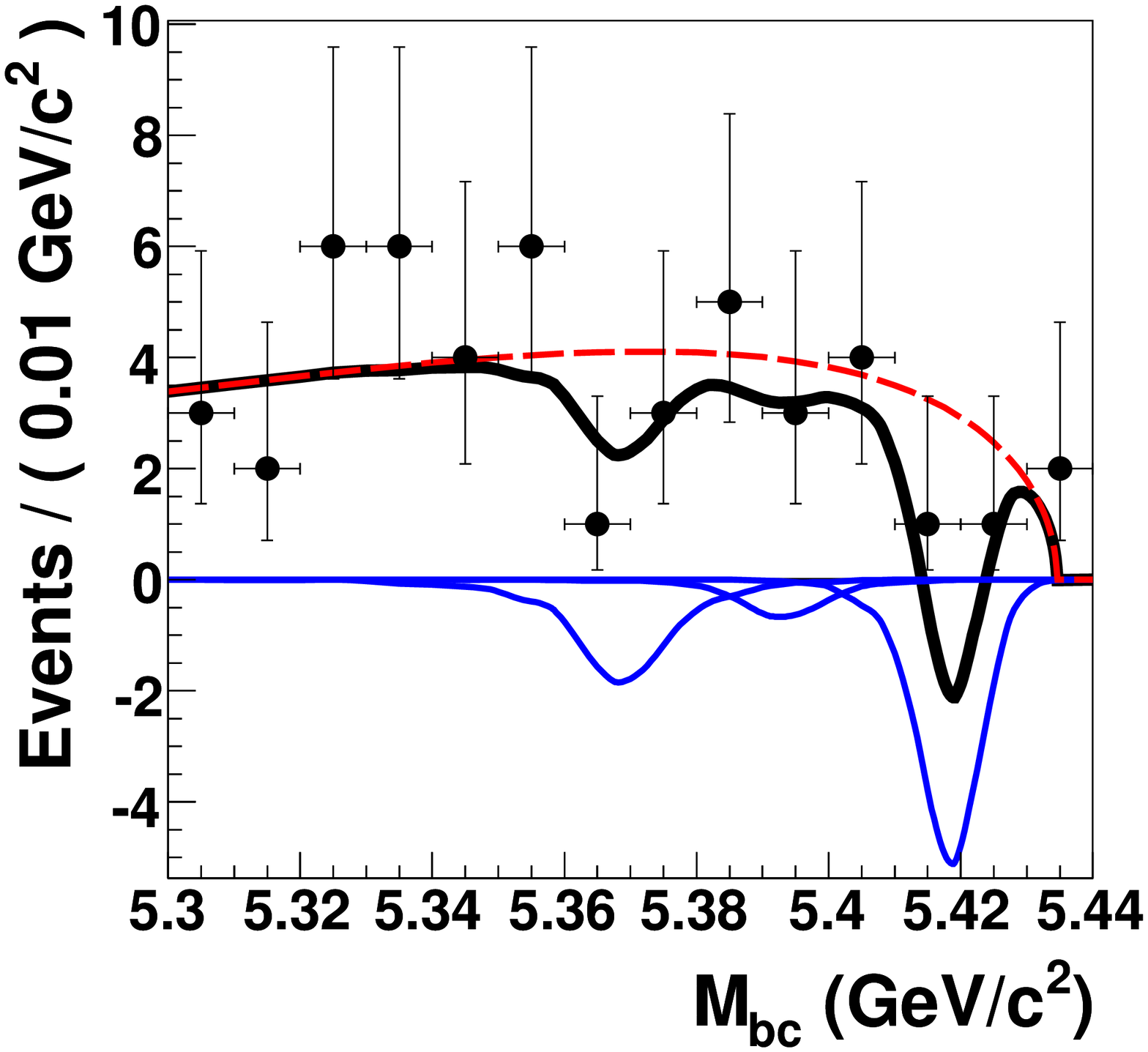} &
   \includegraphics[width=0.23\textwidth]{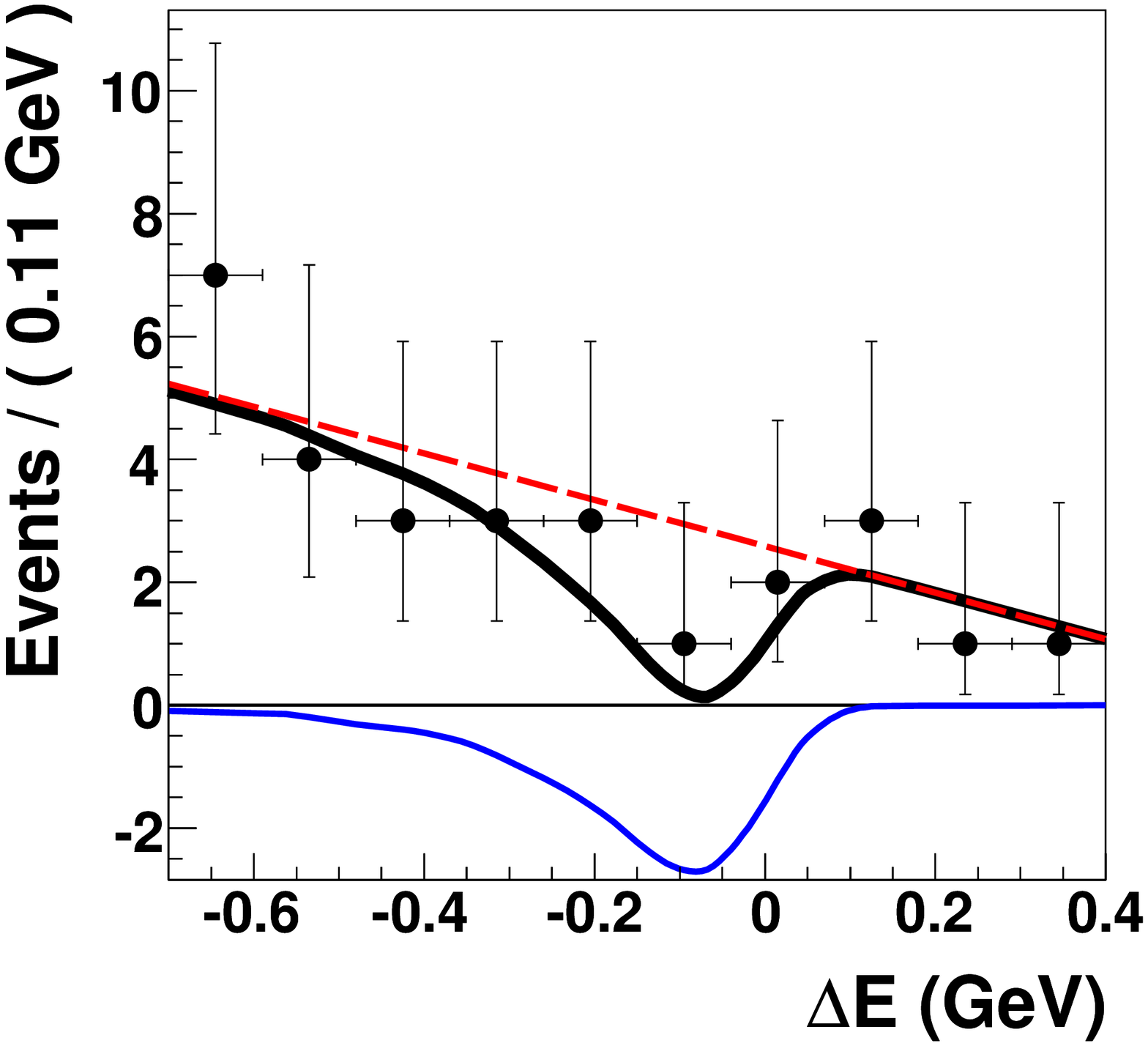} \\ 
 \end{tabular}
\caption{\mbc\ and \deltae\ projections together with fit results for the \Bstogg\ mode. 
The points with error bars represent data, the thick solid curves are the fit functions, the thin solid curves are the signal functions, and the dashed curves show the continuum contribution. On the \mbc\ figure, signals from \BsBs, \BsstBs\ and \BsstBsst\ appear from left to right. On the \deltae\ figure, due to the requirement $\mbc > 5.4$ \gevct\ only the \BsstBsst\ signal contributes.}
\label{figure:data_bs2gg}
\end{figure}

Systematic uncertainties are listed in Table~\ref{table:systerrorsensitivity}. The error on the signal reconstruction efficiency is dominated by uncertainty on the efficiency of the \SFW\ requirement.  This uncertainty is evaluated by comparing efficiencies in data and MC using the \BstoDspi\ control sample. 
For the \Bstophig\ mode, we take as systematic uncertainty the \BF\ difference between the results of the nominal fit and the results of a fit where the continuum is parametrized with a second-order polynomial function for \deltae. For the \Bstogg\ mode, the limit obtained with the nominal continuum parametrization is found to be conservative.
For the \Bstophig\ mode, systematic uncertainties on \BF\ are evaluated by repeating the fit with each parameter successively varied by plus or minus one standard deviation around its central value. The positive and negative uncertainty in \BF\ are obtained from the quadratic sum of the corresponding deviations from the \BF\ value returned by the nominal fit. The significance of the branching fraction measurement is defined as $\sqrt{2 (\ln{\mathcal{L}_{\rm max}-\ln{\mathcal{L}_{0}}})}$, where $\mathcal{L}_{\rm max}$ is the likelihood returned by the nominal fit and $\mathcal{L}_{0}$ is the likelihood returned by the fit with \BF\ set to zero. Systematic uncertainties are included by choosing the lowest significance value returned by the fits used to evaluate the systematic uncertainty. The \Bstophieta\ background is the only source of systematic uncertainty having a non-negligible effect on the significance. For the \Bstogg\ mode, the 90\% CL limit, $\BF_{\rm limit}$, is computed by likelihood integration, according to $\int_{0}^{\BF_{\rm limit}} \mathcal{L}(\BF)\,d\BF = 0.9 \times \int_{0}^{1} \mathcal{L}(\BF)\,d\BF$. Systematic uncertainties are included by convolving the likelihood function with Gaussian distributions for the parameters giving rise to systematic uncertainty.

\renewcommand{\arraystretch}{1.2}
\begin{table}
\centering
\caption{Systematic uncertainties.}
\begin{tabular}{l c c }\hline
Source                    & \Bstophig               & \Bstogg                 \\ \hline
Photon reconstruction efficiency & $2.2\%$                 & $2 \times 2.2\%$        \\ 
Tracking efficiency              & $2 \times 1\%$          & --                      \\ 
Kaon identification efficiency   & $2 \times 1.1\%$        & --                      \\  
$\SFW$ requirement efficiency    & $10\%$                  & $10\%$                  \\ 
MC statistics                    & $0.8\%$                 & $1.1\%$                 \\ \cline{2-3}
$\epsilon$ (quadratic sum)       & $10.7\%$                & $11.0\%$                \\ \hline 

Signal shape                     & $^{+3.2}_{-4.2}\%$        & negl.\                  \\ \hline
\deltae\ continuum shape         & $^{+2.5}_{-0.0}\%$        & negl.\                  \\ \hline 
$\Bs$ backgrounds                & $^{+0.0}_{-1.2}\%$        & negl.\                      \\ \hline

$\BF(\phitoKK)$                  & $1.2\%$                 & --                      \\ \hline

\Lint                            & \multicolumn{2}{c}{$1.4\%$} \\
\sigmabb                         & \multicolumn{2}{c}{$5.0\%$} \\         
$f_s$                            & \multicolumn{2}{c}{$^{+15}_{-12}\%$} \\ \cline{2-3}
\NBs (quadratic sum)             & \multicolumn{2}{c}{$^{+16}_{-13}\%$} \\ \hline 

\fBsstBsst                       & \multicolumn{2}{c}{$^{+7.5}_{-9.7}\%$} \\ \hline \hline

Total (quadratic sum)            & $^{+21}_{-20}\%$  & $^{+21}_{-19}\%$ \\ \hline

\end{tabular}
\label{table:systerrorsensitivity}
\end{table}

In summary, we observe for the first time a radiative penguin decay of the \Bs\ meson in the \Bstophig\ mode.  We measure $\BF(\Bstophig) = \bfBstophigss$, which is in agreement with both the SM predictions and with extrapolations from measured \BptoKstg\ and \BztoKstg\ decay branching fractions. No significant signal is observed in the \Bstogg\ mode and we set an upper limit at the 90\% CL of $\BF(\Bstogg) < \limitbfBstogg$.  This limit significantly improves on the previously reported one and is only an order of magnitude larger than the SM prediction, providing the possibility of observing this decay at a future Super $B$-factory~\cite{superbelle,superb}.

We thank the KEKB group for excellent operation of the
accelerator, the KEK cryogenics group for efficient solenoid
operations, and the KEK computer group and
the NII for valuable computing and Super-SINET network
support.  We acknowledge support from MEXT and JSPS (Japan);
ARC and DEST (Australia); NSFC (China); 
DST (India); MOEHRD, KOSEF and KRF (Korea); 
KBN (Poland); MES and RFAAE (Russia); ARRS (Slovenia); SNSF (Switzerland); 
NSC and MOE (Taiwan); and DOE (USA).


\begin{thebibliography}{99}
\bibitem{chargeconj}
The inclusion of the charge-conjugate mode is implied throughout this letter.
\end{thebibliography}
\end{document}